\documentclass[prl,twocolumn,showpacs,preprintnumbers,amsmath,amssymb,floatfix,superscriptaddress]{revtex4}

\usepackage{graphicx,float}
\usepackage{subfigure}
\usepackage{dcolumn}
\usepackage{bm}
\usepackage{mathrsfs}
\usepackage{color}

\begin{document}

\title{Sub-wavelength position measurements with running wave driving fields}

\author{J\"{o}rg \surname{Evers}}
\affiliation{Max-Planck-Institut f\"{u}r Kernphysik, Saupfercheckweg 1,
D-69117 Heidelberg, Germany}

\author{Sajid \surname{Qamar}}
\affiliation{Max-Planck-Institut f\"{u}r Kernphysik, Saupfercheckweg 1,
D-69117 Heidelberg, Germany}
\affiliation{Centre for Quantum Physics, COMSATS Institute of Information Technology, Islamabad, Pakistan}

\date{\today}

\begin{abstract}
A scheme for sub-wavelength position measurements of quantum particles is discussed, which operates with running-wave laser fields as opposed to standing wave fields proposed in previous setups. The position is encoded in the phase of the applied fields rather than in the spatially modulated intensity of a standing wave. Therefore, disadvantages of standing wave schemes such as cases where the atom remains undetected since it is at a node of the standing wave field are avoided. Reversing the directions of parts of the driving
laser fields allows to switch between different magnification levels, and thus  to optimize the localization.
\end{abstract}


\maketitle

The localization of quantum particles is an intriguing area of research that already in the early days of quantum mechanics led to much discussion and an improved understanding of the underlying theory. Further research is fueled by the need for effective structuring and measuring schemes at small length scales in many modern applications. Approaches involving light, however, due to diffraction are typically restricted to an accuracy of order of the involved wavelength $\lambda$~\cite{bornwolf}.
This limit could be overcome in techniques operating at the sub-wavelength scale which are based on near-field techniques~\cite{near} or rely on distinguishable quantum objects~\cite{distinguish}. Alternatively, sub-wavelength measurements in the optical far field have been suggested. A particular class of quantum optical localization schemes suitable to determine the position of a quantum particle on a sub-wavelength scale makes use of standing wave driving fields~\cite{Walls,Rempe,Rempe-2,Herkommer,review,dual-meas,sajid-2,paspalakis-1,kapale-1,kapale-2,dark-res,Evers}. These on the one hand act as a ruler for the position measurement, and on the other hand encode position information into the atomic dynamics via their position dependent intensity. 

Standing wave based schemes, however, typically do not work equally well over the whole range of potential positions throughout one wavelength. In the worst case, the atom is located at a node of the standing wave field, such that no direct detection is possible. 
Another disadvantage arises in recent schemes with improved localization that facilitate more than one position-dependent fields with different frequencies~\cite{Evers,Luling}. This frequency difference leads to a beat, and  the relative light field intensities are not periodic in space on a length scale of a typical wavelength $\lambda$. Thus, the localization is not periodic with $\lambda$, which so far has been neglected in the theoretical analysis.
Third, current standing-wave based schemes usually require an additional classical measurement to determine one out of several potential positions due to the periodicity of the standing wave. 

In this Letter, we discuss sub-wavelength position measurement of quantum particles using running wave fields, which allows to circumvent the problems associated with standing wave fields. The position is encoded in the phases of the applied fields rather than in the spatially modulated intensity of a standing wave. Therefore, cases where the atom remains undetected since it is at a node of the standing wave field are avoided. The phase sensitivity arises since the driving fields are applied such that they form a closed interaction loop~\cite{closed-loop,morigi,mahmoudi,experiment}. The number of applied laser fields determines the maximum resolution of the measurement schemes, which can be improved by adding more fields. The higher resolution, however, comes at the cost of more potential positions per laser wavelength. But this limitation can be avoided since we find that for a given number of laser fields, changing the directions of individual driving laser fields allows to switch between different magnification levels. Thus, the position can first be determined on a coarser scale with few potential positions, and then be refined using a higher magnification.

\begin{figure}[b]
\centering
\includegraphics[width=8cm]{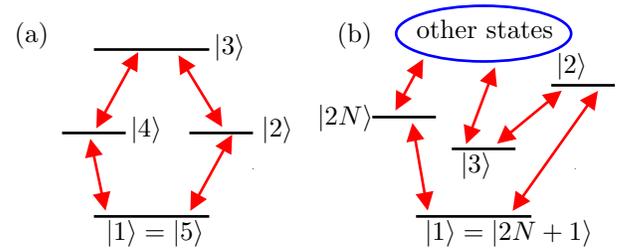}
\caption{\label{fig-system}(Color online) (a) The diamond scheme with
$2N=4$ states that is considered as specific example. (b) Generic atomic loop system
with $2N$ states. Red arrows indicate driving laser fields, spontaneous emission is not depicted.}
\end{figure}

We start by considering a four-level system in diamond configuration
as a basic atomic level setup suitable
for our localization scheme, see Fig.~\ref{fig-system}(a)~\cite{morigi}. In the final part, we will 
generalize our analysis to general closed-loop systems as indicated in
Fig.~\ref{fig-system}(b).
The diamond scheme consists of one ground state $|1\rangle$, two intermediate 
states $|2\rangle$ and  $|4\rangle$, and one excited state $|3\rangle$. 
All four electric-dipole allowed transitions $|1\rangle-|2\rangle$, $|2\rangle-|3\rangle$,
$|4\rangle-|3\rangle$ and $|1\rangle-|4\rangle$ are driven by coherent 
laser fields. 
Thus, starting from the state with lowest energy $|1\rangle$, the system can evolve 
in a non-trivial loop sequence of laser interactions via $|2\rangle$, $|3\rangle$ and
$|4\rangle$  back to the initial state. We denote the atomic levels with state indices increasing 
along the closed loop path from 1 to $2N$, and identify state $|1\rangle$ with $|2N+1\rangle$ for the sake of
simpler analytical expressions. The spontaneous decay rates from level $|i\rangle$ 
to the levels $|j\rangle$  are denoted as $2 \gamma_{ji}$. 
The Hamiltonian in dipole and rotating wave approximation 
is given by
\begin{align}
H =& \sum_{j=1}^{4} \, E_j \, A_{jj} 
+ \hbar( g_{21}\,A_{21}e^{i\alpha_{21}}+
+ g_{32}\,A_{32}e^{i\alpha_{32}}\nonumber\\
&+ g_{34}\,A_{34}e^{i\alpha_{34}}
+ g_{41}\,A_{41}e^{i\alpha_{41}} + H.c. )\,,
\end{align}
where $A_{ij}=|i\rangle\langle j|$,  the moduli of the Rabi frequencies are $g_{ij}$, 
and the energies of the involved states are denoted by
$E_j$ ($j\in\{1,\dots,4\}$).
The parameters $\alpha_{ij}=\omega_{ij}t - \vec{k}_{ij}\vec{r}+\phi_{ij}$
contain  $\omega_{ij}$ as the laser frequencies,  the
wave vectors $\vec{k}_{ij}$, absolute phases $\phi_{ij}$  arising from both the
laser field and the dipole moment, and $\vec{r}$ as the position
of the atom. We further define the 
transition frequencies $\bar{\omega}_{ij}=(E_i-E_j)/\hbar$
and laser field detunings $\Delta_{ij} = \omega_{ij}-\bar{\omega}_{ij}$.
In a suitable interaction picture, the dynamics of the system density matrix $\rho$ is determined by $\partial_t \rho = -i [V,\rho] /\hbar + \mathcal{L}\rho$, where the Liouvillian $\mathcal{L}$ describes spontaneous emission and the Hamiltonian $V$ is 
\begin{align}
V =& - \hbar \Delta_{21} A_{22}
- \hbar (\Delta_{21}+\Delta_{32})  A_{33}
- \hbar \Delta_{41}  A_{44}
\nonumber \\
 +& \hbar (g_{21}   A_{21}  + g_{34}   A_{34} +
g_{41}   A_{41} + g_{32}  e^{-i\Phi}\, A_{32} + H.c.)\,.
\nonumber 
\end{align}
As it is well known for closed-loop systems, the dynamics depends on
the phase~\cite{morigi,mahmoudi}
\begin{align}
\label{phi}
\Phi &= \Delta t-\vec{K}\vec{r}+\phi_0 \,, 
\end{align}
where $\Delta = \sum_{i=1}^{2N} \Delta_{i+1,i}$
is known as the multiphoton detuning,
$\vec{K} = \sum_{i=1}^{2N} \vec{k}_{i+1,i}$ as the
wave vector mismatch,
and $\phi_0 = \sum_{i=1}^{2N} \phi_{i+1,i}$ as the relative phase
of the involved driving fields. Here, $X_{ij}=-X_{ji}$ for 
$X\in\{\Delta, \vec{k}, \phi\}$.

In order to proceed with the analysis, we make certain assumptions on 
the setup and on the level structure. First, we adjust all laser fields
to be on resonance, i.e., $\Delta = 0$. Further, we note that the 
phase $\phi_0$ is a relative phase which can be controlled independent
of the position of the atom by phase-locking the different laser fields
to each other~\cite{experiment}.  
Next, to fix a definite setup, we assume that all transitions couple to circularly polarized light with polarization vector in the $x$-$y$-plane and propagation directions of the laser fields parallel to the $z$-axis,
\begin{subequations}
\begin{align}
\vec{k}_{21} &= k_{21}\,\hat{e}_z\,, &
\vec{k}_{34} &= \epsilon_{34} \, k_{34}\,\hat{e}_z \,, \\
\vec{k}_{32} &= k_{32}\,\hat{e}_z\,, &
\vec{k}_{41} &= \epsilon_{41} \, k_{41}\,\hat{e}_z \,.
\end{align}
\end{subequations}
Here, $\epsilon_{ij} \in \{-1,1\}$ and $\hat{e}_z$ is the unit vector in $z$ direction.
\begin{figure}[t]
\centering
\includegraphics[width=8cm]{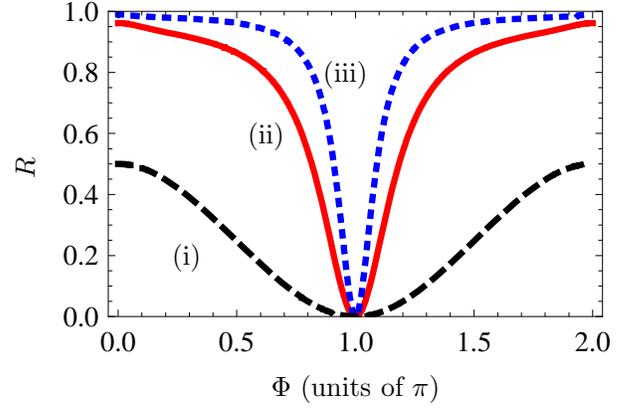}
\caption{\label{fig-ratio}(Color online) Ratio $R$ of the
fluorescence intensities on transitions $|3\rangle \to
|2\rangle$ and $|2\rangle \to |1\rangle$. (i) $\Omega = \gamma$,
(ii) $\Omega = 5\gamma$, (iii) $\Omega = 10 \gamma$.}
\end{figure}
Then, the closed-loop phase $\Phi$ evaluates to
\begin{align}\label{final-1}
\Phi = (k_{21}+k_{32}- \epsilon_{34} k_{34}-\epsilon_{41} k_{41})\,z + \phi_0\,.
\end{align}
Eq.~(\ref{final-1}) is the origin of our central results. It demonstrates that
it is possible to choose a laser setup such that the multiphoton detuning
does not contribute to the closed-loop phase, whereas the dependence
on the wave vectors leads to a position dependence of the phase $\Phi$ controllable
by the propagation directions of the laser fields. In order to analyze
this control further, we approximate $k_{ij}\approx k = 2\pi/\lambda$, and
obtain 
\begin{align}
\label{phi-2}
\Phi = 2\pi  \, \frac {\xi}{\lambda} \, z + \phi_0\,,
\end{align}
where $\xi = (2-\epsilon_{34}-\epsilon_{41})$. From Eq.~(\ref{phi-2}), we find that the closed loop phase $\Phi$ changes by $2\pi$ over a position range of $\lambda/\xi$.
It is interesting to note that in closed-loop systems studied before, typically the phase-matching condition $\vec{K}\vec{r}\approx 0$ was assumed in order to avoid a position dependence on a sub-wavelength scale. In contrast, for sub-wavelength localization, this dependence is exactly what is required.

In order to make use of Eq.~(\ref{final-1}) in a realistic setup, an 
easily accessible observable is required which depends on the closed-loop phase $\Phi$.
A number of potential observables such as state populations, fluorescence spectra, or light
propagation dynamics have been identified in the literature~\cite{closed-loop,morigi,mahmoudi,experiment}. In the following,
we choose the fluorescence intensity of light emitted on the different transitions
as the simplest observable accessible in the optical far field. The fluorescence
intensity on the different transitions is proportional to the spontaneous
decay rate on the transition times the population in the upper state of the
respective transition. 

\begin{figure}[t]
\includegraphics[width=8cm]{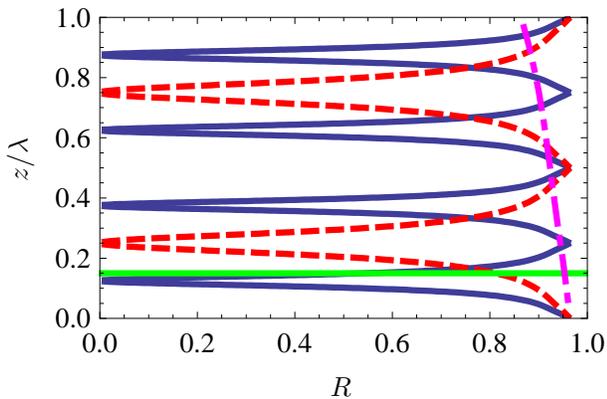}
\caption{\label{fig-pos}(Color online) Potential positions $z$ depending
on the measured ratio $R$. Solid blue lines show $\xi=4$, dashed red lines indicate $\xi=2$, and the dash-dotted magenta line shows $\xi=0.25$. The horizontal green line indicates the assumed particle position $z=0.15 \lambda$.}
\end{figure}
Since the intensities also depend on experimental conditions such as the 
distance of the detector to the atom, the area of the detector
or its efficiency, it is convenient to consider the ratio of two intensities as the observable in order to reduce the dependence on these quantities.
We define $R$  as the ratio of the fluorescence intensity on transition $|3\rangle \to
|2\rangle$ to the intensity on transition $|2\rangle \to
|1\rangle$. 

The measurement then could proceed as follows. After applying the driving fields,
the intensities of the light emitted on the two transitions is measured.
The light from the different transitions can be distinguished via the 
polarizations or the frequencies. From the ratio $R$ of the two intensities, the phase $\Phi$ can be determined. Via Eqs.~(\ref{final-1}) or (\ref{phi-2}), $\Phi$ is related to the position $z$ of the atom.
More specific, assuming for the sake of simpler analytical results
all spontaneous decay rates to be equal to $\gamma$, and all Rabi frequencies
equal to $x\, \gamma$ with $x \in \mathbb{R}$, we obtain $R = N(x)/D(x)$, where
\begin{subequations}\label{ratio}
\begin{align}
N(x) &=2 x^2 \cos(\Phi/2)^2 [-(3+2 x^2)^2+4 x^4 \cos\Phi]\,,\\
D(x) &= -18-51 x^2-28 x^4-2 x^6+\nonumber \\
&\qquad x^2 (9+4 x^2) \cos\Phi+2 x^6 \cos(2 \Phi)\,.
\end{align}
\end{subequations}
\begin{figure}[t]
\centering
\includegraphics[width=8cm]{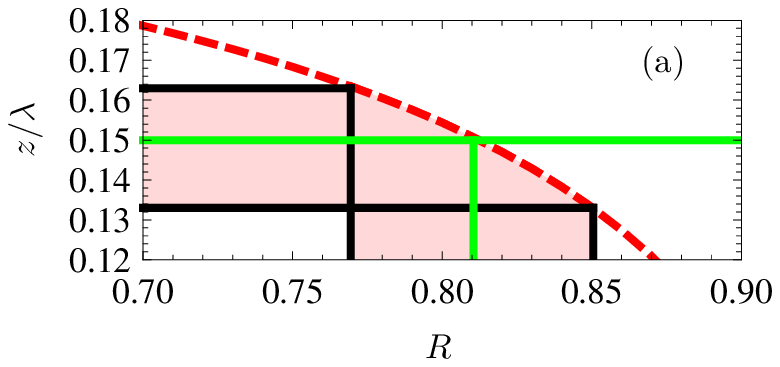}
\includegraphics[width=8cm]{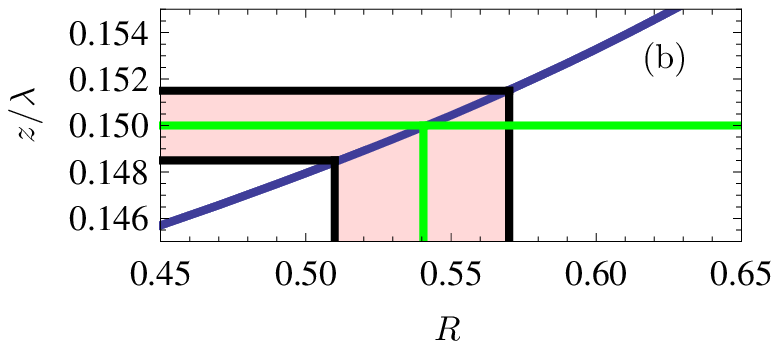}
\includegraphics[width=8cm]{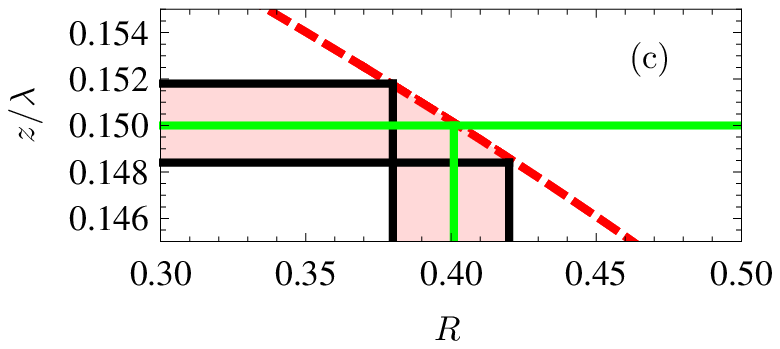}
\caption{\label{fig-pos2}(Color online) Particle position $z$ depending
on the measured ratio $R$, with parameters as in Fig.~\ref{fig-pos}. The shaded area indicates uncertainties due to an imperfect measurement of $R$. The horizontal green line indicates the assumed particle position $z=0.15 \lambda$. (a) Laser configuration $\xi=2$, relative phase $\phi_0=0$; (b) Laser configuration $\xi=4$, relative phase $\phi_0=0$; (c) Laser configuration $\xi=2$, relative phase $\phi_0=\pi/4$. Note the different position axis scale in (a).}
\end{figure}
This ratio $R$ is plotted for different driving field strengths $\Omega=x\gamma$ in Fig.~\ref{fig-ratio}.
Ideally, the ratio should be chosen such that
there is a high slope $|\partial R/\partial \Phi|$ throughout the
whole range of phases $\Phi$. From Fig.~\ref{fig-ratio}, it can be seen 
that $x\approx 5$ offers a good compromise of high maximum slope and 
low minimum slope over the whole range of $0\leq \Phi\leq 2\pi$.
Then, the relative phase can be determined well from the measured
intensity ratio. 

Setting $x=5$, we now proceed to extract potential positions of the atom 
from the measured fluorescence intensity ratio $R$. For this, we calculate
the positions leading to the measured rations $R$ using Eqs.~(\ref{phi-2}) and
(\ref{ratio}). 

The results for $\xi\in\{2,4\}$ are shown in Fig.~\ref{fig-pos}.
Like in standing wave localization schemes, each ratio $R$ corresponds to several potential positions $\{z_i\}$~\cite{Evers}. As suggested previously, an additional classical measurement
could be used to determine the true position out of several potential
positions~\cite{Rempe-2,sajid-2,Evers}. 
However, the dash-dotted magenta curve in Fig.~\ref{fig-pos} suggests how this classical
measurement can be avoided. For this, $\xi\ll 1$ is required.
This can be achieved with $\epsilon_{34}=\epsilon_{41}=1$ in Eq.~(\ref{final-1}) if the 
magnitudes of the four wave vectors slightly differ due to unequal transition frequencies. 
Alternatively, a small mismatch
of the propagation directions from $\pm \hat{e}_z$ allows to tune $\xi$ to small non-zero values. 
In the latter case, however, the driving fields acquire polarization components that may drive unwanted 
transitions. Small values of $\xi$ allow to  measure the approximate position of the particle 
on a coarse scale, because the phase $\Phi$ then changes on a scale $\lambda/\xi \gg \lambda$. 
Using this method, the need for an additional classical measurement
common to the localization schemes suggested so far is eliminated.

We now turn to a discussion of measurements on a sub-wavelength scale with higher $\xi$.
In the following, we assume as a concrete example that a particle is located at position $z=0.15\,\lambda$. This position is indicated as solid green line in Figs.~\ref{fig-pos} and \ref{fig-pos2}.
We further assume that for $\xi=2$, a measurement of the ratio $R$ obtained a value of $R=0.81\cdot (1\pm0.05)$ with an overall uncertainty of $5\%$. Finally, we assume that the correct branch is identified by a measurement with small $\xi$ determining the appxoximate position of the particle. Fig.~\ref{fig-pos2}(a) then demonstrates how the result for $R$ together with its overall error is translated into position information. It can be seen that the slope of the curve essentially determines the uncertainty in the position measurement. In Fig.~\ref{fig-pos2}(a), a localization up to an uncertainty of about 20\% is achieved. 

This uncertainty can be reduced using two methods. First, changing the propagation direction the laser beams and thus $\xi$ leads to an improvement of the measurement. Fig.~\ref{fig-pos2}(b) shows the example for $\xi=4$. Again, an overall error of 5\% in the measurement of $R$ is assumed. It can be seen that due to the smaller slope than in the case $\xi=2$, the position error is reduced by about one order of magnitude to about 2\%.
Thus, by reversing the direction of one of the laser fields, the localization is greatly improved. 
Second, the relative phase $\phi_0$ in Eq.~(\ref{phi}) can be used to improve the measurement. It turns out that a change of $\phi_0$ shifts all curves in Fig.~\ref{fig-pos} along the position axis. Thus the phase $\phi_0$ can be optimized such that the true position of the particle leads to a measured ratio $R$ in a range where the slope of the curves in Fig.~\ref{fig-pos} is small. This is demonstrated in Fig.~\ref{fig-pos2}(c) for $\xi=2$. In this figure, a relative phase $\phi_0 = \pi/4$ was used to shift the ratio $R$ corresponding to the true position from about $R=0.81$ in Fig.~\ref{fig-pos2}(a) to about $R=0.4$ in Fig.~\ref{fig-pos2}(c). At this ratio, the slope $|\partial z/\partial R|$ is much smaller, and this reduces the position error by almost one order of magnitude to about 2.5\%. In general, the best localization results are obtained if the optimization via the phase $\phi_0$ is combined with laser configurations with large $\xi$.

From Eq.~(\ref{final-1}), it is clear that the slope $|\partial z/\partial R|$ of the position against the ratio $R$ is decreased by increasing $\xi$, such that errors in $R$ lead to smaller uncertainties in $z$. But as in previous standing-wave based schemes, an improvement of the localization increases the number of potential positions~\cite{Rempe-2,sajid-2,Evers}. 
Thus it is useful to first determine the position approximately using a small $\xi$, and then switch to higher prefactors with better resolution around the known approximate position. Due to this analogy to conventional optical microscopy, the directions of the laser fields can be interpreted  as determining the magnification of the localization measurement.

In the final part, we generalize our results to an extended loop
system. For $2N$ states labeled with increasing state index along
the loop path as before, the generalized loop phase for the case of resonant fields becomes
\begin{align}
\Phi = \phi_0 + z\, \sum_{i=1}^{2N} \epsilon_{i+1,i}\, k_{i+1,i}\,,
\end{align}
where $\epsilon_{i+1,i}\in\{-1,1\}$ determine the propagation directions
of the corresponding laser beams. Approximating again $k_{ij}\approx 2\pi/\lambda$, one obtains
\begin{subequations}
\begin{align}
\Phi &\approx 2\pi \, \frac{\xi}{\lambda}\, z + \phi_0\,,\\
\xi &\in \{0,2,\dots,2N\}\,.
\end{align}
\end{subequations}
The value of $\xi$ can be controlled by the choice of the propagation directions
$\epsilon_{i,j}$. Thus we find that the prefactor $\xi$ already present in Eq.~(\ref{phi-2}) that enables one to determine the magnification of the localization scheme can be chosen in a wide range. Since for a given $\xi$, the relative phase $\Phi$ changes by $2\pi$ over a position range of $\lambda/\xi$, a more extended closed loop scheme enables one to achieve a better resolution for the position determination, and to switch between more
magnification levels. 
Potential restrictions arise from the requirement that the individual transitions should ideally be addressed individually by laser fields. This, however, is not a strict requirement, as long as the multiphoton resonance condition is fulfilled, but restricts the possible settings for $\xi$. 
In particular in more extended systems, other observables than the simple ratio of two fluorescence intensities can be expected to lead to further improvement for the position determination.


\begin{thebibliography}{99}

\bibitem{bornwolf}M. Born and E. Wolf, {\it{Principles of Optics}} (Cambridge University Press, Cambridge, England, 1999).

\bibitem{near}A. Lewis {\it et al.}, 
Nature Biotechnology {\bf 21}, 1378 (2003).

\bibitem{distinguish}E. Betzig, Opt. Lett. {\bf 20}, 237 (1995).




\bibitem{Walls} P. Storey, M. Collett, and D. F. Walls, Phys. Rev. Lett. {\bf 68}, 472 (1992); Phys. Rev. A {\bf 47}, 405 (1993); R. Quadt, M. Collett, and D. F. Walls, Phys. Rev. Lett. {\bf 74}, 351 (1995).
%
\bibitem{Rempe}S. Kunze, K. Dieckmann, and G. Rempe, Phys. Rev. Lett. {\bf 78}, 2038 (1997).
%
\bibitem{Rempe-2} F. Le Kien {\it et al.}, 
Phys. Rev. A {\bf 56}, 2972 (1997).
%

\bibitem{Herkommer} A. M. Herkommer, W. P. Schleich, and M. S. Zubairy, J. Mod. Opt. {\bf 44}, 2507 (1997).
%

\bibitem{review}J. E. Thomas and L. J. Wang, Phys. Rep. {\bf 262}, 311 (1995).


\bibitem{dual-meas} H. Nha {\it et al.}, 
Phys. Rev. A {\bf 65}, 033827 (2002).
%

\bibitem{sajid-2} S. Qamar, S.-Y. Zhu, and M. S. Zubairy, Phys. Rev. A. {\bf 61}, 063806 (2000).
%


\bibitem{paspalakis-1} E. Paspalakis and P. L. Knight, Phys. Rev. A {\bf 63}, 065802 (2001).
%

\bibitem{kapale-1} M. Sahrai {\it et al.}, 
Phys. Rev. A {\bf 72}, 013820 (2005); K. T. Kapale and M. S. Zubairy, {\it ibid.} {\bf 73}, 023813 (2006).
%

\bibitem{kapale-2} G. S. Agarwal and K. T. Kapale, J. Phys. B: At. Mol. Opt. Phys. {\bf 39}, 3437 (2006).
%

\bibitem{dark-res} C. Liu {\it et al.}, 
Phys. Rev. A {\bf 73}, 025801 (2006).
%


\bibitem{Evers} J. Evers, S. Qamar, and M. S. Zubairy, Phys. Rev. A {\bf 75}, 053809 (2007).
%
\bibitem{Luling}L. Jin {\it et al.}, J. Phys. B: At. Mol. Opt. Phys. {\bf 41}, 085508 (2008).

\bibitem{closed-loop}
S. J. Buckle {\it et al.}, 
Opt. Acta {\bf 33}, 2473 (1986); 
D. V. Kosachiov, B. G. Matisov, and Y. V. Rozhdestvensky, 
J. Phys. B {\bf 25}, 2473 (1992).


\bibitem{morigi}G. Morigi, S. Franke-Arnold, and G.-L. Oppo, Phys. Rev. A {\bf 66}, 053409 (2002); S. Kajari-Schr\"oder {\it et al.}, 
{\it ibid.} {\bf 75}, 013816 (2007).

\bibitem{mahmoudi}M. Mahmoudi and J. Evers, Phys. Rev. A {\bf 74}, 063827 (2006)

\bibitem{experiment}E. A. Korsunsky {\it et al.},
Phys. Rev. A {\bf 59}, 2302 (1999); 
H. Kang {\it et al.}, {\it ibid.}  {\bf 73}, 011802(R) (2006).

\end{thebibliography}
\end{document}